\documentclass[pra,preprint,showpacs]{revtex4}

\begin{document}
%\draft
\title{The spinor field theory of the photon}
\author{Ruo Peng WANG\footnotemark
\footnotetext{email: rpwang@pku.edu.cn}}
\affiliation{ Physics Department, Peking University, Beijing 100871,
P.R.China }
\date{\today}

%\narrowtext
\begin{abstract}

I introduce a spinor field theory for the photon. The
three-dimensional vector electromagnetic field and the
four-dimensional vector potential are components of this spinor
photon field. A spinor equation for the photon field is derived from
Maxwell's equations,the relations between the electromagnetic field
and the four-dimensional vector potential, and the Lorentz gauge
condition. The covariant quantization of free photon field is done,
and only transverse photons are obtained. The vacuum energy
divergence does not occur in this theory. A covariant ``positive
frequency'' condition is introduced for separating the photon field
from its complex conjugate in the presence of the electric current
and charge.

\end{abstract}

\pacs{03.70.+k, 11.10.-z,03.65.Pm}
%\showpacs
\maketitle
\section{Introduction}

The electromagnetic interaction is the best studied one among the
four known fundamental interactions. In the frame of quantum theory,
the photon is the quantum of the electromagnetic field. There are
several ways to represent the electromagnetic field: by the
three-dimensional electric field and magnetic field vectors, by the
$4\times 4$ electromagnetic tensor, or by the four-dimensional
potential vector \cite{ld, jks}. The electric and magnetic fields
are directly related to the energy density of the electromagnetic
field. The four-vector potential is directly related to the
Lagrangian density of interaction. But however, none of these fields
can be regarded as the photon field, because the photon density can
not be expressed as inner products of these fields with their
adjoint fields. In this paper, I introduce the spinor photon field
that satisfies a spinor equation similar to the Dirac equation for
the electron. The three-dimensional vector electric field, the
three-dimensional magnetic vector field and the four-dimensional
vector potential are components of this spinor photon field. The
spinor equation for the photon field is based on the Maxwell
equations, the relations between the electromagnetic field and the
four-dimensional vector potential, and the Lorentz gauge condition.
. The Lagrangian densities for the free photon field and for the
photon field in interaction with the matter are established.
Covariant quantization of photon field is carried out, and only
transverse photons emerge from the quantization procedure. The
vacuum state of the photon field is found to have null energy. The
solution for the photon field in the presence of the electric
current and charge is found, and a covariant ``positive frequency''
condition is introduced for separating the photon field from its
complex conjugate.

The Maxwell equations, the relations between the electromagnetic
field and the four-dimensional vector potential, the Lorentz gauge
condition are rewritten as two eight-component spinor equations
 in Sec.\ref{sec:ema}. The spinor photon field is introduced
 in Sec. \ref{sec:photon}. The quantization of the photon field is
 treated in the Sec. \ref{sec:quan}, and the photon field in the
 presence of electric current and charge is analyzed in Sec. \ref{sec:int} .

\section{The spinor equation for the electromagnetic field}
\label{sec:ema}

Maxwell's equations for the electric and magnetic fields
$\vec E$ and $\vec H$ in the presence of a charge
density $\rho$ and a current density $\vec j$ can be written in the following form:
\begin{eqnarray}\label{max1}
        && \frac{\partial }{\partial x_0}(\sqrt{\varepsilon_0}
        \vec{E})
    = \nabla \times (\sqrt{\mu_0} \vec{H}) - \sqrt{\mu_0} \vec{j} \\
    \label{max2}
       && \frac{\partial}{\partial x_0}(\sqrt{\mu_0} \vec{H})
    =-\nabla \times (\sqrt{\varepsilon_0} \vec{E})\\ \label{max3}
       && 0=-\nabla \cdot (\sqrt{\mu_0} \vec{H})  \\ \label{max4}
       && 0=\nabla \cdot (\sqrt{\varepsilon_0} \vec{E})
    -\sqrt{\varepsilon_0} \rho
\end{eqnarray}
where $\varepsilon_0$ is the vacuum permittivity, $\mu_0$ is the
magnetic permeability of the vacuum, and $x_0=c t$. Because all the
electric field, the magnetic field, the current density and the
charge density are real quantities, they are completely described by
their positive frequency components. An alternative way for writing
the above equations is to introduce an eight components spinor
electromagnetic field and an eight components spinor electric
current density defined by
\begin{equation}
    \psi_{em}(x)=(
        \sqrt{\varepsilon_0}E_{1}(x)\;\;
        \sqrt{\varepsilon_0}E_{2}(x)\;\;
        \sqrt{\varepsilon_0}E_{3}(x)\;\;0\;\;
        \sqrt{\mu_0}H_{1}(x)\;\;
        \sqrt{\mu_0}H_{2}(x)\;\;
        \sqrt{\mu_0}H_{3}(x)\;\;0
        )^T,
\end{equation}
and
\begin{equation}
    j_e(x)= \sqrt{\mu_0}\bigl(j_1(x) \;\;j_2(x)\;\;j_{3}(x)\;\;0\;\;
        0 \;\; 0 \;\; 0 \;\;j_0(x) \bigr)^T,
\end{equation}
where $j(x)=(c\rho(x), \vec j(x))$ is the four-vector current density.

The Maxwell equations (\ref{max1}-\ref{max4}) now can be written as
\begin{equation}\label{sem}
    \frac{\partial}{\partial x_0} \psi_{em}(x)=
    -\vec{\alpha}_e \cdot \nabla \psi_{em}-j_e(x)
\end{equation}
where
\begin{equation}
    \alpha_{e1}=\left(\begin{array}{cccc}
        0 & 0 &  0 & -i\sigma_2 \\
        0 & 0 & -i\sigma_2 & 0 \\
        0 & i\sigma_2 &  0 & 0 \\
          i\sigma_2 & 0 &  0 & 0
        \end{array}\right)\;,\;
    \alpha_{e2}=\left(\begin{array}{cccc}
        0 & 0  & 0 & -I_2 \\
        0 & 0  & I_2 & 0 \\
          0 & I_2  & 0 & 0 \\
          -I_2 & 0 & 0 & 0
        \end{array}\right)\;,\;
    \alpha_{e3}=\left(\begin{array}{cccc}
        0 & 0 &  i\sigma_2 & 0 \\
        0 &  0 &  0 & -i\sigma_2 \\
          -i\sigma_2 &  0 &  0 & 0 \\
            0 & i\sigma_2 & 0 & 0
        \end{array}\right)\;,
\end{equation}
with
\begin{equation}
    \sigma_2=\left(\begin{array}{cc}
        0 & -i  \\
        i & 0
    \end{array}\right)\;
    \mbox{and}\;
    I_2=\left(\begin{array}{cc}
        1 & 0  \\
        0 & 1
    \end{array}\right).
\end{equation}
We have
\begin{equation}
    \alpha_{em} \cdot \alpha_{en} + \alpha_{en} \cdot \alpha_{em}
    = 2 \delta_{nm} \;, \; n,m = 1, 2, 3.
\end{equation}

The electromagnetic field can be described by the four-vector
potential $A(x)=(\phi(x)/c, \vec A(x))$. The relation between the
electric and magnetic fields and the four-vector potential, and the
Lorentz gauge condition can be written as:
\begin{eqnarray}
      \frac{\partial}{\partial x_0}
    \bigl( \sqrt{\varepsilon_0} \vec{A} \bigr) &=&  - \nabla
    \bigl(\sqrt{\varepsilon_0} A_0\Bigr)
    -\frac{1}{c}\sqrt{\varepsilon_0} \vec{E} \label{pot1} \\
      0 &=& \nabla \times \Bigl(\sqrt{\varepsilon_0}\vec{A} \bigr)
    - \frac{1}{c} \sqrt{\mu_0} \vec{H}  \label{pot2} \\
      \frac{\partial}{\partial x_0} \bigl(\sqrt{\varepsilon_0} A_0 \bigr)&=&
     -\nabla \cdot \bigl(\sqrt{\varepsilon_0} \vec{A} \bigr) \label{pot3}
\end{eqnarray}
The relations (\ref{pot1})-(\ref{pot3}) can be rewritten as
\begin{equation}\label{spo}
    \frac{\partial}{\partial x_0}\psi_a(x)  =
    \vec{\alpha}_e \cdot \nabla \psi_a(x)
    - \frac{1}{\hbar c} \psi_{em}(x)
\end{equation}
where spinor potential field $\psi_a(x)$ is defined by
\begin{equation}
    \psi_{a}(x)= \frac{\sqrt{\varepsilon_0}}{\hbar}\bigl(A_1(x)
    \;\;A_2(x)\;\;A_{3}(x)\;\;0\;\;0 \;\; 0 \;\; 0 \;\;A_0(x)
    \bigr)^T.
\end{equation}

One may observe that the equation (\ref{spo}) also holds if we
replace the Lorentz gauge condition with the following one:
\begin{equation}\label{pot4}
    \frac{\partial}{\partial x_0} \bigl(\sqrt{\varepsilon_0} A_0
    \bigr)=
     -\nabla \cdot \bigl(\sqrt{\varepsilon_0} \vec{A} \bigr)
     -\frac{1}{c} \psi_{em8},
\end{equation}
where $\psi_{em8}$ is an arbitrary scalar constant. In this case,
the spinor electromagnetic field has the following form
\begin{equation}
    \psi_{em}(x)=(
        \sqrt{\varepsilon_0}E_{1}(x)\;\;
        \sqrt{\varepsilon_0}E_{2}(x)\;\;
        \sqrt{\varepsilon_0}E_{3}(x)\;\;0\;\;
        \sqrt{\mu_0}H_{1}(x)\;\;
        \sqrt{\mu_0}H_{2}(x)\;\;
        \sqrt{\mu_0}H_{3}(x)\;\;
        \psi_{em8}
        )^T.
\end{equation}

According to properties of the electric field $\vec E$,
magnetic field $\vec H$ and four-vector potential $A(x)$
under continuous space-time transformations,
we have the following relation for $\psi_{em}(x)$
and $\psi_a(x)$ under a Lorentz transformation
\begin{equation}
    \psi^\prime_{em}(x^\prime) = \exp \bigl(-\vec \varphi \cdot \vec l \bigr)
    \psi_{em}(x^\prime),
\end{equation}
and
\begin{equation}
    \psi^\prime_{a}(x^\prime) = \exp \bigl(\vec \varphi \cdot (
    \vec \alpha_e - \vec l) \bigr) \psi_{a}(x^\prime),
\end{equation}
where
\begin{equation}
    \vec \varphi = \frac{\vec v}{v}
    \Bigl( \ln \sqrt{1+\frac{v}{c}}-\ln \sqrt{1-\frac{v}{c}}\Bigr).
\end{equation}
Under a rotation characterized by the rotation angle $\vec \phi $,
expressed as an axial vector, we have
\begin{equation}
    \psi_{em}^\prime(x^\prime) = \exp {( i \vec \phi \cdot \vec s)}
    \psi_{em}(x^\prime)
\end{equation}
and
\begin{equation}
    \psi_{a}^\prime(x^\prime) = \exp {( i \vec \phi \cdot \vec s)}
    \psi_{a}(x^\prime),
\end{equation}
with
\begin{equation}
    \vec s=\left(\begin{array}{cc}
        \vec \Sigma & 0      \\
        0      & \vec \Sigma
        \end{array}\right)\;,\;
    \vec l= \left(\begin{array}{cc}
        0 & i\vec \Sigma      \\
        -i\vec \Sigma & 0
        \end{array}\right),
\end{equation}
and
\begin{equation}
    \Sigma_{1}=\left(\begin{array}{cccc}
        0 & 0 &  0 & 0 \\
        0 & 0 & -i & 0 \\
        0 & i &  0 & 0 \\
            0 & 0 &  0 & 0
        \end{array}\right)\;,\;
    \Sigma_{2}=\left(\begin{array}{cccc}
        0 & 0  & i & 0 \\
        0 & 0  & 0 & 0 \\
           -i & 0  & 0 & 0 \\
            0 & 0  & 0 & 0
        \end{array}\right)\;,\;
    \Sigma_{3}=\left(\begin{array}{cccc}
        0 & -i &  0 & 0 \\
            i &  0 &  0 & 0 \\
            0 &  0 &  0 & 0 \\
            0 &  0 &  0 & 0
        \end{array}\right).
\end{equation}
The following commutation relation holds for $\vec s$:
\begin{equation}\label{scom}
    [s_{n},s_{m}] = i \sum_{p=1}^3 \varepsilon_{nmp}s_{p} .
\end{equation}
Equations (\ref{sem}) and (\ref{spo}) are invariant under continuous
space-time transformations (See the Appendices).

\section{the photon field }
\label{sec:photon}

 One can use $\psi_{em}(x)$ or $\psi_a(x)$ to
represent the electromagnetic field, but it is not possible to
express the photon density as a inner product of $\psi_{em}(x)$ or
$\psi_a(x)$ with its adjoint field. Therefore neither
$\psi_{em}(x)$, nor $\psi_a(x)$ can be regarded as the photon field.
The concept of photon is closely related to monochromatic
electromagnetic plane waves, so we consider a monochromatic plane
wave
\begin{equation}
    \psi_{em}^k(x), \psi_{a}^k(x) \propto \exp (-ikx)
\end{equation}
in absence of electric current and charge. Let $\psi_{em}^{+k}(x)$
and $\psi_{a}^{+k}(x)$ be the positive frequency
parts of $\psi_{em}^k(x)$ and $\psi_{a}^k(x)$.
We find that the inner product $\psi_{em}^{+k \dag}(x) \psi_{em}^{+k}(x)$
is equal to the time average of energy density,
which should equal to, in the case of a monochromatic wave,
a product of the photon density and the photon energy $\hbar k_0 c$.
On other hand, we have
\begin{equation}
    \epsilon_0 \vec E^{+k *}(x) \cdot \vec E^{+k}(x)
    = \mu_0 \vec H^{+k *}(x) \cdot \vec H^{+k}(x)=
    \frac{1}{2} \psi_{em}^{+k \dag}(x) \psi_{em}^{+k}(x),
\end{equation}
and
\begin{equation}
    i \vec E^{+k *}(x) \cdot \vec A^{+k}(x) =
    -i \vec A^{+k *}(x) \cdot \vec E^{+k}(x) = \frac{1}{k_0 c}
    \vec E^{+k *}(x) \cdot \vec E^{+k}(x),
\end{equation}
thus the photon density is equal to
\begin{equation}
    -i \psi_{a}^{+k \dag}(x) \psi_{em}^{+k}(x) + i \psi_{em}^{+k\dag}(x)
    \psi_{a}^{+k}(x).
\end{equation}
We have
\begin{equation}\label{em2f}
    -i \psi_{a}^{+k \dag}(x) \psi_{em}^{+k}(x) + i \psi_{em}^{+k\dag}(x)
    \psi_{a}^{+k}(x)
    =
    \left(\psi_{em}^{+k\dag}(x) \;\;  \psi_{a}^{+k \dag}(x)\right)
    \tau_2
    \left( \begin{array}{c}
    \psi_{em}^{+k}(x) \\ \psi_{a}^{+k \dag}(x)
    \end{array} \right),
\end{equation}
where
\begin{equation}
    \tau_2 = \left( \begin{array}{c c}
    0 & i\beta_e \\ -i\beta_e & 0
    \end{array} \right)
    \mbox{  with  }
    \beta_e = \left( \begin{array}{c c}
    I_4 & 0 \\ 0 & -I_4
    \end{array} \right),
\end{equation}
and $I_4$ is the $4 \times 4$ unit matrix.

Based on the relation (\ref{em2f}), we define the photon field
$\psi_f(x)$ as
\begin{equation}
    \psi_f(x) \equiv  \frac{1}{16\pi^4} \int_{k_0>0} d^4 k
    \exp \bigl(ik(x^\prime-x) \bigr) \left( \begin{array}{c}
    \psi_{em}(x^\prime) \\ \psi_{a}(x^\prime)
    \end{array} \right),
\end{equation}
One may observe that the condition $k_0 > 0$ is covariant for free
photon field, because it does not contain Fourier components with
$k_0 < |\vec k|$. According to Eqs. (\ref{sem}) and (\ref{spo}), the
free photon field satisfies the following equation:
\begin{equation}\label{sph}
    i \hbar \frac{\partial}{\partial x_0} \psi_{f}(x)
    = -i\hbar \vec{\alpha}_w \cdot \nabla \psi_{f}(x)
    - \frac{i}{c}\beta_{-}\psi_{f}(x)
\end{equation}
with
\begin{equation}
    \vec \alpha_{w} = \left( \begin{array}{c c}
    \vec \alpha_e & 0 \\ 0 & -\vec \alpha_e
    \end{array} \right) ,\;\;
    \beta_{-} = \left( \begin{array}{c c}
    0 & 0 \\ I_8 & 0
    \end{array} \right).
\end{equation}
where $I_8$ is the $8 \times 8$ unit matrix.

The invariance of Eq. (\ref{sph}) under continuous space-time transformations is
assured by the invariance of
Eqs. (\ref{sem}) and (\ref{spo}). We have
\begin{equation}\label{lrntz}
    \psi^\prime_{f}(x^\prime) = \exp (\vec \varphi \cdot \vec \Lambda)
    \psi_{f}(x^\prime),
\end{equation}
with
\begin{equation}\label{lmbd}
    \vec \Lambda =
    \left(\begin{array}{cc} - \vec l & 0 \\
    0 & \vec \alpha_e - \vec l
    \end{array}\right)
\end{equation}
for Lorentz transformations, and
\begin{equation}
    \psi_{f}^\prime(x^\prime) = \exp {(i \vec \phi \cdot \vec s_f)} \psi_{f}(x^\prime)
\end{equation}
under space rotations, where
\begin{equation}
    \vec s_f=\left(\begin{array}{cc}
        \vec s & 0      \\
        0      & \vec s
        \end{array}\right).
\end{equation}

Eq. (\ref{sph}) is invariant also under space inversion and time
reversal. It is easy to verify that $\tau_0\psi_f(x_0,-\vec x)$ and
 $\tau_3\psi^*_f(- x_0,\vec x)$ satisfy the same spinor equation
Eq. (\ref{sph}) as $\psi_f( x_0,\vec x)$. Where
\begin{equation}
    \tau_0 = \left( \begin{array}{c c}
    -\beta_e & 0\\ 0 & -\beta_e
    \end{array} \right),\;
    \tau_3 = \left( \begin{array}{c c}
    \beta_e & 0\\ 0 & -\beta_e
    \end{array} \right).
\end{equation}

The equation for the free photon field can be derived from the following Lagrangian density
\begin{equation} \label{lag}
    {\cal L}_0 = i\hbar \bar \psi_f\left( \frac{\partial}{\partial t}
    + c \vec {\alpha}_{w} \cdot \nabla \right) \psi_f
    + i \bar \psi_f \beta_{-}\psi_f,
\end{equation}
where $\bar \psi_f(x) = \psi^\dag_f(x) \tau_2$ is the adjoint field.

One may observe that there are totally 15 component equations for
photon field. Among these 15 equations only 11 equations are
independent, and the other 4 equations (corresponding
Eqs.(\ref{max2}) and (\ref{max3})) are direct conclusion of these 11
equations. By means of variational calculation, all these 11
independent equations can be obtained.

The conjugate field of $\psi_f$ is
\begin{equation}
    \pi_f = \frac {\partial {\cal L}_0}{\partial \dot {\psi_f}}
    = i \hbar \bar \psi_f.
\end{equation}

The Hamiltonian now can be calculated:
\begin{eqnarray}\label{hml}
    H_0 & = &  \int d^3 \vec{x} \left( \pi_f \dot{\psi_f} - {\cal L}_0 \right)
    \nonumber\\
    & = & \int d^3 \vec{x} \bar \psi_f \left( -i\hbar c \vec {\alpha}_{w} \cdot \nabla
    - i \beta_{-}\right) \psi_f.
\end{eqnarray}

The Lagrangian density (\ref{lag}) is invariant under a global phase
change of the photon field $\psi_f(x)$ . This implies the
conservation of the photon number $N$ for free photon field:
\begin{equation}
    N = \int \rho_{ph} d^3 \vec x,
\end{equation}
and
\begin{equation}
    \frac{\partial}{\partial t} \rho_{ph} + \nabla \cdot \vec j_{ph}=0,
\end{equation}
where the photon density $\rho_{ph}(x) $ is given by the inner product between the
photon field $\psi_f(x)$ and
its adjoint $\bar \psi_f (x) $:
\begin{equation}\label{rho}
    \rho_{ph}(x) = \bar \psi_f (x) \psi_f(x)
\end{equation}
and
\begin{equation}
    \vec {j}_{ph}(x) = c \bar \psi_f(x) \vec \alpha_w \psi_f(x)
\end{equation}
is the photon current density. One may observe that the photon
density defined by Eq.(\ref{rho}) may take negative values. But
however, when we talk about photons we refer to electromagnetical
fields with well defined frequencies. By direct calculation, one can
verify that $\rho_{ph} \ge 0$ if $\psi_f(x)$ has a well defined
frequency.

According to the relation between symmetries and conservation laws
\cite{noether, mandl}, we may obtain the following expressions for
the momentum $\vec P$ and the angular momentum $\vec M$ of the free
photon field:
\begin{equation}
    \vec {P} = - i \hbar \int d^3 \vec{x} \bar \psi_f\nabla \psi_f,
\end{equation}
and
\begin{equation}
    \vec {M} =  \int d^3 \vec{x} \bar \psi_f [\vec{x} \times
    (- i\hbar \nabla)] \psi_f + \int d^3 \vec{x} \bar \psi_f
    (\hbar \vec{s}_f) \psi_f.
\end{equation}
It is clear that $\vec s_f$ can be interpreted as the spin operator of the photon field.
According to expression (\ref{scom}), we have
\begin{equation}
    [s_{fn},s_{fm}] = i \sum_{p=1}^3 \varepsilon_{nmp}s_{fp}\;.
\end{equation}

\section{quantization of the photon field}
\label{sec:quan}

It is convenient to quantize the photon field in the momentum space.
To do this, we have to find firstly the plan wave solutions of the
photon field. By substituting the following form of solution
\begin{equation}\label{pln}
    \psi_f(x) \propto \exp{(-i k x)}w{(\vec{k})}
\end{equation}
into the spinor equation (\ref{sph}), we find
\begin{equation}\label{eqw}
    \left( \vec{\alpha_w} \cdot \vec{k} - k_0 -
    \frac{i \beta_{-}}{\hbar c} \right)
     w{(\vec{k})}= 0.
\end{equation}
Eq. (\ref{eqw}) permits two independent nontrial solutions with $k_0
= | \vec{k} | $. They can be chosen as
\begin{eqnarray}\label{sol}
    w_{\pm 1}(\vec k) &=& \frac{1}{2\sqrt{\hbar c}}
    \Bigl(\hbar c |\vec k| (q_1 \pm i r_1) \;\; \hbar c |\vec k|
    (q_2 \pm i r_2) \;\;
    \hbar c |\vec k| (q _3 \pm i r_3) \;\; 0    \;\;
    \nonumber \\ &&
    \hbar c |\vec k|
    (r_1 \mp i q_1) \; \;
    \hbar c |\vec k| (r_2 \mp i q_2) \;\;
    \hbar c |\vec k| (r_3 \mp iq_3) \;\; 0 \;\;
    \nonumber \\ &&
    -i q_1 \pm r_1 \;\; -i q_2 \pm r_2 \;\;
    -i q_3 \pm r_3 \;\;0 \;\; 0 \;\; 0 \;\; 0 \;\; 0 \Bigr)^T,
\end{eqnarray}
where $\hat k = \vec k/|\vec k|$, and $\vec q$ and $\vec r$ are two vectors of unity
satisfying the following conditions:
\begin{equation}
    \hat k \times \vec q = \vec r,\;\;
    \hat k \times \vec r = -\vec q,\; \;
    \vec q \times \vec r =\hat k,\;\; \mbox {and}\;\;
    \vec r(-\hat k) = - \vec r(\hat k).
\end{equation}
$ w_{+1}(\vec k) $ and $ w_{-1}(\vec k)$  are orthogonal:
\begin{equation}
    \bar w_h (\vec{k})w_{h^\prime}(\vec{k}) = w_h^\dag
    (\vec{k})\tau_2
    w_{h^\prime}(\vec{k}) = \delta_{hh^\prime} |\vec k| .
\end{equation}
We also have
\begin{equation}
    \bigl( \hat{k} \cdot \vec s \bigr) w_{h}(\vec k)
    = h w_{h}(\vec k),\; h=\pm1,
\end{equation}
and
\begin{equation}
    \vec s_f ^2 w_{h}(\vec k)
    =s(s+1)w_{h}(\vec k)=2w_{h}(\vec k),
\end{equation}
therefore photons are of spin s=1.

Having plan wave solutions of the photon field $\psi_f(x)$, we may
now expand $\psi_f(x)$ in plane waves
\begin{equation}\label{exp}
    \psi_f(x)=\sum_{\vec k} \sum_{h} \frac{1}{\sqrt {V |\vec k|}}
    e^{-ikx}w_h{(\vec{k})} b_h(\vec k)  \;,
    \bar \psi_f(x)=\sum_{\vec k} \sum_{h} \frac{1}{\sqrt {V|\vec k|}}
    e^{ikx} \bar w_h {(\vec{k})} b^\dag _h(\vec k)
\end{equation}
with $k_0 = |\vec k|$.
According to relations (\ref{lag}) and
(\ref{exp}), the Lagrangian of the photon field can expressed as a
function of the variables $ q_{h\vec k} (t) $:
\begin{equation} \label{lgf}
    L_0(t,q) = \sum_{\vec k} \sum_{h} \hbar q^\dag_{h\vec k}(t)
    (i\frac{\partial}{\partial t}   - c |\vec k| ) q_{h\vec k} (t),
\end{equation}
with
\begin{equation}
    q_{h\vec k}( t) = b_h(\vec k) \exp{(-i \omega t)},\;\omega = c k_0.
\end{equation}

The conjugate momentum of $q_{h\vec k}(t)$ can be calculated, and we
have
\begin{equation}
    p_{h\vec k}(t) = \frac {\partial L_0}{\partial \dot q_{h\vec k}( t)}
    = i \hbar b^\dag_h(\vec k) \exp{(i \omega t)}.
\end{equation}
By applying the quantization condition $ [ q_{h\vec k} , p_{h^\prime
\vec k^\prime} ] = i \hbar \delta_{hh^\prime} \delta_{\vec k \vec
k^\prime}$ we find the following commutation relation for $
b_{\pm1}(\vec k)$ and $ b^\dag _{\pm 1}(\vec k)$
\begin{equation}\label{quant}
    [b_h(\vec k), b^\dag_{h^\prime}(\vec k^\prime)]=
    \delta_{hh^\prime}\delta_{\vec k \vec k^\prime}\;.
\end{equation}
$ b_{\pm 1}(\vec k)$ and $ b^\dag _{\pm 1}(\vec k)$
are just the photon annihilation operator and the photon
creation operator. The Hamiltonian of the photon field can also be calculated. We obtain
\begin{equation}
    H_0 = \sum_{\vec k} \sum_{h} p_{h\vec k}\dot q_{h\vec k} - L_0
    = \sum_{\vec k} \sum_{h} \hbar \omega b^\dag_h(\vec k)b_h(\vec k).
\end{equation}
We observe that the vacuum energy of the photon field is zero.

The commutation relations for the photon field can be written in a
covariant form. According to the commutation
relations (\ref{quant}) and the expression (\ref{exp}), we have
\begin{equation}\label{quantc}
    [ \psi_{fl}^\dag (x), \psi_{fm}(x^\prime)]= D_{lm}(x-x^\prime),
    \;\;    \text{with }\; l,m =1,2,\cdots,8 ,
\end{equation}
where the $8\times 8$ matrix $D(x)$ is given by the following
expression
\begin{equation}
    D(x) = \frac{\hbar c}{2(2\pi)^3}\int_{k_0 > 0}
    d^4 k \delta (k^2)\left[k_0 \vec k \cdot \vec l
    + (\vec k \cdot \vec l) (\vec k \cdot \vec l)\right]
    e^{-ikx}.
\end{equation}
The replacement
\begin{equation}
    \frac{1}{V}\sum_{\vec k} \longrightarrow
    \frac{1}{(2\pi)^3}\int d^3 \vec k
\end{equation}
was used in obtaining the relation (\ref{quantc}). Under Lorentz transformations,
$D(x)$ transforms to
\begin{equation}
    D^\prime (x^\prime) = \exp \bigl(-\vec \varphi \cdot \vec l \bigr)
    D(x^\prime) \exp \bigl(-\vec \varphi \cdot \vec l \bigr).
\end{equation}
One can verify with no difficulty that by using the expansions
(\ref{exp}), the commutation relations (\ref{quant}) can be derived
from commutation relation (\ref{quantc}). Therefore, the commutation
relations (\ref{quant}) and (\ref{quantc}) are equivalent.

\section{ interaction between photon field and matter }
\label{sec:int}

In the presence of electric current and charge, according to Eqs.
(\ref{sem}) and (\ref{spo}), we have the following equation for the
photon field:
\begin{equation}\label{sfi}
    i \hbar \frac{\partial}{\partial x_0} \psi_{f}(x) = \bigl(-i\hbar   \vec{\alpha}_{w} \cdot \nabla
    - \frac{i}{c}\beta_{-} \bigr) \psi_{f}(x)
    -i\hbar J_f(x),
\end{equation}
where spinor current density $J_f(x)$ is given by
\begin{equation}
    J_f(x) = \left( \begin{array}{c} j^+_e(x) \\ 0 \end{array}
    \right),
\end{equation}
and $j^+_e(x)$ satisfies the relation
\begin{equation}
    j^+_e(x) + j^{+*}_e(x)=j_e(x).
\end{equation}
It is would be nature to request that spinor current density
$j^+_e(x)$ to have only positive frequency Fourier components. But
the spinor current density $j_e(x)$ may contain Fourier components
with $|\vec k| > |k_0|$, and for these Fourier components this
separation does hold for all frames of reference. In other words,
the positive frequency condition is not covariant. A covariant form
of this condition can be written as: $ k p > 0 $, where $ p $ is a
well defined 4-vector. For each physical system, there always is a
well defined 4-vector, namely the total energy-momentum 4-vector of
the system. therefore, we have
\begin{equation}
    \int d^4 x e^{ikx}J_f(x) = \theta(kp) \int d^4 x
    e^{ikx} \left( \begin{array}{c} j_e(x) \\ 0 \end{array}
    \right),
\end{equation}
with
\begin{equation}
    \theta(x)=\left\{\begin{array}{cll}
                1 & {\mbox if } & x > 0 \\
                1/2 & {\mbox if } & x=0 \\
                0 & {\mbox if } & x<0
                \end{array}\right . ,
\end{equation}
and $ p = (p_0, \vec p)$ the total energy-momentum 4-vector of the
photon field and the charged matter field under consideration. One
may observe that, in the ``center of mass '' frame in which the
total momentum $ \vec p $ of the whole system in interaction is
null, $j_e^+(x)$ have only positive frequency Fourier components.

It is easy to verify that the equation (\ref{sfi}) can be derived
from the following Lagrangian density
\begin{equation}\label{lgdi_1}
    {\cal L}(x) = {\cal L}_{0}(x)+{\cal L}^f_{int}(x)
\end{equation}
with the Lagrangian density of interaction given by
\begin{equation}
     {\cal L}^f_{int}(x) =  i\hbar c\bigl[\psi_f^\dag(x)
    \tau_2 J_f(x) - J^\dag_f(x) \tau_2 \psi_f(x)\bigr].
\end{equation}
According to the definition of $J_f(x)$, we have
\begin{equation}
    \int_{-\infty}^{\infty}  dx_0
    \int_V d^3 \vec x
    \bigl[\psi_f^\dag(x)
    \tau_2 J^*_f(x) - J^T_f(x) \tau_2 \psi_f(x)\bigr]=0.
\end{equation}
Therefore it is not necessary to separate $J_f(x)$ from $J^*_f(x)$
in the Lagrangian density of interaction, and the equation
(\ref{sfi}) can also be derived from the following Lagrangian
density
\begin{equation}\label{lgdi_2}
    {\cal L}(x) = {\cal L}_{0}(x)+{\cal L}_{int}(x)
\end{equation}
with the Lagrangian density of interaction given by
\begin{equation}
     {\cal L}_{int}(x) =  i\hbar c\bigl[\psi_f^\dag(x)
    \tau_2 J_s(x) - J^\dag_s(x) \tau_2 \psi_f(x)\bigr],
\end{equation}
with
\begin{equation}
     J_s(x) = J_f(x) + J^*_f(x).
\end{equation}
According to the relation between the photon field $\psi_f(x)$ and
the four-vector potential $A_f(x)$, the Lagrangian density of
interaction ${\cal L}_{int}(x)$ can also be written as
\begin{equation}
     {\cal L}_{int}(x) = - A^\dag_f(x) j(x) - j(x) A_f(x),
\end{equation}
where $A_f(x)$ is the `` positive frequency'' part of $ A(x)$:
\begin{equation}
    \int d^4 x e^{ikx}A_f(x) = \theta(kp) \int d^4 x
    e^{ikx} A(x).
\end{equation}

One may observe that if $j(x)$ commutes with $A_f(x)$ and
$A^\dag_f(x)$, the Lagrangian density of interaction ${\cal
L}_{int}(x)$ would becomes $ - j(x)A(x) $, as in the classical
electrodynamics.

The equation (\ref{sfi}) can be solved. We have
\begin{equation}\label{slf1}
    \psi_f(x) = \psi^0_f(x) + \int d^4 x^\prime G_f(x-x^\prime) J_f(x^\prime),
\end{equation}
where $\psi^0_f(x)$ is the free photon field given by expressions (\ref{exp}),
and $G_f(x)$ is the Green function for
the photon field:
\begin{equation}
    G_f(x) = \frac{1}{16\pi^4} \int d^4 k \frac{ -i \exp (-ikx )}
    {k_0^2 - |\vec k|^2 +i\epsilon } \left(k_0+\vec \alpha_w \cdot \vec k
    -i\frac{\beta_-}{\hbar c}\right).
\end{equation}

\section{Conclusion}

I introduced a spinor field theory for the photon. The spinor
equation for the photon field is equivalent to Maxwell's equations
together with the relations between the four-vector potential and
electric and magnetic fields, and the Lorentz gauge condition for
the 4-vector potential. The quantization of free photon field is
done, and only transverse photons are obtained. The vacuum energy
divergence does not occur in this theory. The solution for the
photon field in the presence of the electric current and charge is
found, and a covariant ``positive frequency'' condition is
introduced for separating the photon field from its complex
conjugate.

\appendix
\section{Invariance of spinor equations for electromagnetic field and potential under Lorentz transformations}

By direct verification, one may find the following relations for matrices $\vec s$ and $\vec l$:
\begin{equation}
    [s_{n},\alpha_{em}]= i \sum_{p=1}^3 \varepsilon_{nmp}
    \alpha_{ep}     \;,\;   n,m=1,2,3,
\end{equation}
and
\begin{equation} \label{lrel}
    \alpha_{en} l_m \alpha_{en} = l_m -(1-\delta_{nm})\alpha_{em}
    \;,\; n,m=1,2,3.
\end{equation}

Let's consider a Lorentz transformation
\begin{equation}
    \left(\begin{array}{c}x_1^\prime\\x_2^\prime
    \\x_3^\prime\\x_0^\prime \end{array} \right)
     = \left(\begin{array}{cccc}
    \cosh\varphi & 0 & 0 & - \sinh\varphi \\
    0 & 1 & 0 & 0 \\
    0 & 0 & 1 & 0 \\
    - \sinh\varphi & 0 & 0 & \cosh\varphi \end{array} \right)
    \left(\begin{array}{c}x_1\\x_2\\x_3\\x_0 \end{array}
    \right).
\end{equation}
We have
\begin{equation}\label{cotr}
    \frac{\partial}{\partial x_1} =
    \cosh\varphi \frac{\partial}{\partial x_1^\prime} - \sinh\varphi
    \frac{\partial}{\partial x_0^\prime} \;,\;
    \frac{\partial}{\partial x_0} =
    \cosh\varphi \frac{\partial}{\partial x_0^\prime} - \sinh\varphi
    \frac{\partial}{\partial x_1^\prime}
\end{equation}
and
\begin{equation}
    j_{1}(x)= \cosh\varphi j_{1}^\prime(x^\prime) +
    \sinh\varphi c\rho^\prime(x^\prime) \;,\;
    c\rho(x)= \cosh\varphi c\rho^\prime(x^\prime) +
    \sinh\varphi j_{1}^\prime(x^\prime).
\end{equation}
The last relations can be written in the terms of $ j_{e}(x)$ and terms $ j_{e}^{\prime}(x^\prime)$:
\begin{equation}
    j_{e}(x) = \exp \bigl(\varphi (l_1 - \alpha_{e1} ) \bigr)   j_{e}^{\prime}(x^\prime).
\end{equation}
By using the relations (\ref{cotr}), the equation (\ref{sem}) can be written then as the following:
\begin{equation}
    (\cosh\varphi - \sinh\varphi \alpha_{e1})
    \frac{\partial} {\partial x_0^\prime }\psi_{em}(x^\prime)
    +\Bigl[(\cosh\varphi \alpha_{e1} - \sinh\varphi)\frac{\partial}
    {\partial x_1^\prime}
    +\alpha_{e2} \frac{\partial}{\partial x_2^\prime}
    +\alpha_{e3} \frac{\partial}{\partial x_3^\prime}\Bigr]
      \psi_{em}(x^\prime) + j_{e}(x^\prime) =0.
\end{equation}
Because
\begin{equation}
    \exp (-\varphi \alpha_{e1}) = \cosh\varphi - \sinh\varphi
    \alpha_{e1},
\end{equation}
so
\begin{eqnarray}
    \exp (-\varphi \alpha_{e1})
    \frac{\partial }{\partial x_0^\prime }\psi_{em} (x^\prime) &=&
    -\Bigl[\exp (-\varphi \alpha_{e1})\alpha_{e1}
    \frac{\partial}{\partial x_1^\prime}
    +
    \Bigl(  \alpha_{e2}\frac{\partial}{\partial x_2^\prime}
    +\alpha_{e3}\frac{\partial}{\partial x_3^\prime} \Bigr)
    \Bigr] \psi_{em}(x^\prime)
        \nonumber \\
    &&
    -\exp \Bigl(\varphi (l_1 - \alpha_{e1}) \Bigr) j^\prime_e (x^\prime).
\end{eqnarray}
But
\begin{equation}
    l_1 \alpha_{e1} = \alpha_{e1} l_1
\end{equation}
then
\begin{eqnarray}\label{emtr}
    \frac{\partial }{\partial x_0^\prime }
    \exp \bigl( -\varphi l_1 \bigr)\psi_{em} (x^\prime) &=&
    -\Bigl[\alpha_{e1}
    \frac{\partial}{\partial x_1^\prime}
    + \exp \Bigl(- \varphi (l_1 - \alpha_{e1}) \Bigr)
    \Bigl(  \alpha_{e2}\frac{\partial}{\partial x_2^\prime}
    +\alpha_{e3}\frac{\partial}{\partial x_3^\prime} \Bigr)
    \exp \bigl( \varphi l_1 \bigr) \Bigr]
        \nonumber \\
    &&
    \exp \bigl( -\varphi l_1 \bigr)\psi_{em}(x^\prime)
    - j^\prime_e (x^\prime).
\end{eqnarray}
According to the relation (\ref{lrel}), we have
\begin{equation}
    \alpha_{em} l_1^n  = (l_1 - \alpha_{e1})^n\alpha_{em}
    \;,\; m=2,3,
\end{equation}
therefore
\begin{equation}
    \alpha_{em} \exp \bigl( \varphi l_1 \bigr) =
    \exp \Bigl( \varphi (l_1 - \alpha_{e1}) \Bigr)\alpha_{em}
    \;,\;
    \exp \bigl( \varphi l_1 \bigr) \alpha_{em}  =
    \alpha_{em}  \exp \Bigl( \varphi (l_1 - \alpha_{e1}) \Bigr)
    \;,\; m=2,3.
\end{equation}
and the equation (\ref{emtr}) becomes
\begin{equation}\label{sem1}
    \frac{\partial }{\partial x_0^\prime} \psi^\prime_{em}(x^\prime)
    = -\vec{\alpha_e} \cdot \nabla \psi^\prime_{em}(x^\prime)
     -j_e^{\prime} (x^\prime),
\end{equation}
with
\begin{equation}
    \psi^\prime_{em}(x^\prime)=\exp \bigl( -\varphi l_1 \bigr)\psi_{em}
    (x^\prime).
\end{equation}

The equation (\ref{sem1}) in the new reference frame has exactly the
same form as Eq. (\ref{sem}), that means the spinor equation for the
electromagnetic field is invariant under Lorentz transformations.
The invariance of Eq. (\ref{spo}) can be shown in a similar way. We
have
\begin{eqnarray}\label{potr}
    &&\exp(\varphi \alpha_{e1})
    \frac{\partial} {\partial x_0^\prime }
    \psi_{a}(x^\prime)
    \\ \nonumber
    &&= \Bigl[\alpha_{e1} \exp(\varphi \alpha_{e1})\frac{\partial}
    {\partial x_1^\prime}
            +\alpha_{e2} \frac{\partial}{\partial x_2^\prime}
    +\alpha_{e3} \frac{\partial}{\partial x_3^\prime}\Bigr]
    \psi_{a}(x^\prime) - \frac{1}{\hbar c}
    \exp \bigl( \varphi l_1 \bigr)\psi^{\prime}_{em}(x).
\end{eqnarray}
But
\begin{equation}
    \exp \bigl( - \varphi l_1 \bigr) \alpha_{em}  =
    \alpha_{em}  \exp \Bigl( \varphi ( \alpha_{e1} - l_1) \Bigr)
    \;,\; m=2,3,
\end{equation}
then Eq. (\ref{potr}) can be reduced to
\begin{equation}
    \frac{\partial} {\partial x_0^\prime }\psi^{\prime}_{a}(x^\prime)
    = \vec{\alpha_e} \cdot \nabla
    \psi^{\prime}_{a}(x^\prime) - \frac{1}{\hbar c}
    \psi^{\prime}_{em}(x),
\end{equation}
with
\begin{equation}
    \psi^\prime_{a}(x^\prime)=\exp \Bigl( \varphi ( \alpha_{e1} - l_1) \Bigr)
    \psi_{a} (x^\prime).
\end{equation}
This equation has exactly the same form as Eq. (\ref{spo}).

\section{Invariance of spinor equations for electromagnetic field and potential under space rotation}

Let's consider an infinitesimal space rotation
\begin{equation}
    x_0^\prime=x_0,\;\;
    x_n^\prime = x_n - \sum_{m,p=1}^{3} \varepsilon_{nmp} \delta_m x_p
    \;\;n=1,2,3.
\end{equation}
We have
\begin{equation}
    \frac{\partial}{\partial x_n} =
    \frac{\partial}{\partial x_n^\prime} -
    \sum_{m,p=1}^{3} \varepsilon_{pmn} \delta_m
    \frac{\partial}{\partial x_p^\prime} \;,\;
    n=1,2,3,
\end{equation}
and
\begin{equation}
    \rho(x^\prime)=\rho^\prime(x^\prime),\;\;
    j_n(x^\prime) = j^\prime_n (x^\prime)
    + \sum_{m,p=1}^{3} \varepsilon_{nmp} \delta_m j^\prime_p(x^\prime)
    \;\;n=1,2,3.
\end{equation}
The last relation is equivalent to
\begin{equation}
    j_{e}(x^\prime) = (1 - i \vec \delta \cdot \vec s )
    j_{e}^{\prime}(x^\prime).
\end{equation}
The equation (\ref{sem}) can be written as
\begin{equation}\label{semrot}
    (1 + i \vec \delta \cdot \vec s)
    \frac{\partial}{\partial x_0^\prime}\psi_{em}(x^\prime) =
    -(1 + i \vec \delta \cdot \vec s)
    \sum_{n=1}^{3} \Bigl[\alpha_{en}
    +\sum_{l,m=1}^{3} \varepsilon_{nlm} \delta_l
    \alpha_{em} \Bigr]
    \frac{\partial}{\partial x_n^\prime} \psi_{em}(x^\prime) -  j_{e}^{\prime}(x^\prime).
\end{equation}
But
\begin{equation}
    \sum_{m=1}^3 \varepsilon_{nlm} \alpha_{em} =
    i s_{l}\alpha_{en}- i \alpha_{en} s_{l},
\end{equation}
so
\begin{equation}
    \sum_{l,m=1}^3 \varepsilon_{nlm} \delta_l \alpha_{em} =
    i \vec \delta \cdot \vec s \alpha_{en} - i \alpha_{en}
    \vec \delta \cdot \vec s .
\end{equation}
Eq. (\ref{semrot}) becomes then
\begin{equation}\label{semrot1}
        \frac{\partial}{\partial x_0^\prime}
    (1 + i \vec \delta \cdot \vec s)\psi_{em}(x^\prime) =
    -\sum_{n=1}^{3} \alpha_{en}
    \frac{\partial}{\partial x_n^\prime}
    (1 + i \vec \delta \cdot \vec s)\psi_{em}(x^\prime) -   j_{e}^{\prime}(x^\prime).
\end{equation}
On other hand, we have $\psi_{em}^\prime(x^\prime) = (1 + i \vec
\delta \cdot \vec s) \psi_{em}(x^\prime)$, so we rewrite Eq.
(\ref{semrot1}) as
\begin{equation}
    \frac{\partial}{\partial x_0^\prime}\psi_{em}^\prime(x^\prime) =
    - \vec \alpha_{e} \cdot \nabla \psi_{em}^\prime(x^\prime) -  j_{e}^{\prime}(x^\prime),
\end{equation}
which has exactly the same form as Eq. (\ref{sem}). So the spinor
equation for the electromagnetic field is invariant under space
rotations. The invariance of Eq. (\ref{spo}) can be demonstrated
exactly in the same way.

\end{document}